\documentclass[sigconf]{acmart}

\usepackage{booktabs}
\usepackage{textcomp}

\usepackage{xspace}

\newcommand{\download}{\url{https://gitlab.com/learnERC/filo}\xspace}
\newcommand{\video}{\url{https://youtu.be/WDvkKj-wnlQ}\xspace}

\AtBeginDocument{%
  \providecommand\BibTeX{{%
    \normalfont B\kern-0.5em{\scshape i\kern-0.25em b}\kern-0.8em\TeX}}}

\copyrightyear{2020} 
\acmYear{2020} 
\acmConference[ASE '20]{35th IEEE/ACM International Conference on Automated Software Engineering}{September 21--25, 2020}{Virtual Event, Australia}
\acmBooktitle{35th IEEE/ACM International Conference on Automated Software Engineering (ASE '20), September 21--25, 2020, Virtual Event, Australia}

\begin{document}

\title{FILO: FIx-LOcus Localization for Backward Incompatibilities Caused by Android Framework Upgrades}


%
\author{Marco Mobilio}
\affiliation{%
  \institution{Department of Informatics, Systems and Communication \\ University of Milano-Bicocca}
  \streetaddress{Viale Sarca, 336}
  \city{Milan}
  \state{Italy}
  \postcode{20126}
}
\email{marco.mobilio@unimib.it}

\author{Oliviero Riganelli}
\affiliation{%
  \institution{Department of Informatics, Systems and Communication \\ University of Milano-Bicocca}
  \streetaddress{Viale Sarca, 336}
  \city{Milan}
  \state{Italy}
  \postcode{20126}
}
\email{oliviero.riganelli@unimib.it}

\author{Daniela Micucci}
\affiliation{%
\institution{Department of Informatics, Systems and Communication \\ University of Milano-Bicocca}
  \streetaddress{Viale Sarca, 336}
  \city{Milan}
  \state{Italy}
  \postcode{20126}
}
\email{daniela.micucci@unimib.it}

\author{Leonardo Mariani}
\affiliation{%
\institution{Department of Informatics, Systems and Communication \\ University of Milano-Bicocca}
  \streetaddress{Viale Sarca, 336}
  \city{Milan}
  \state{Italy}
  \postcode{20126}
}
\email{leonardo.mariani@unimib.it}

\newcommand{\marco}[1]{\textcolor{orange}{{\it [Marco says: #1]}}}

\renewcommand{\shorttitle}{FILO}

\begin{abstract}
Mobile operating systems evolve quickly, frequently updating the APIs that app developers use to build their apps. Unfortunately, API updates do not always guarantee backward compatibility, causing apps to not longer work properly or even crash when running with an updated system. This paper presents FILO, a tool that assists Android developers in resolving backward compatibility issues introduced by API upgrades. FILO both suggests the method that needs to be modified in the app in order to adapt the app to an upgraded API, and reports key symptoms observed in the failed execution to facilitate the fixing activity. Results obtained with the analysis of 12 actual upgrade problems and the feedback produced by early tool adopters show that FILO can practically support Android developers. 
FILO can be downloaded from \download, and its video demonstration is available at \video.
\end{abstract}



\maketitle
\section{Introduction}
\label{sec:introduction}
Mobile operating systems evolve rapidly to satisfy emerging user needs, to fix bugs, and to 
exploit the most recent hardware and software upgrades~\cite{McDonnell:APIStability:ICSM:2013,AndroidVersionHistory}. For example, the Android operating system is updated on average every two months~\cite{AndroidVersionHistory}. Such frequent updates affect the correctness of existing apps~\cite{Wei:AndroidFragmentation:ASE:2016,Li:DeprecatedAPI:MSR:2018}, causing developers to struggle to maintain their applications usable with the most recent versions of the operating systems~\cite{Linares-Vasquez:StackOverflowDiscussions:ICPC:2014}.

Many of these issues are caused by the evolution of the Application Programming Interfaces (APIs) provided by the Android framework to control the interaction between the apps and the operating system. For instance, Wei et al.~\cite{Wei:AndroidFragmentation:ASE:2016} found that more than one third of the backward compatibility issues affecting the most popular Android apps are due to the evolution of the APIs; while Mostafa et al.~\cite{Mostafa:BackwardIncompatibilities:ISSTA:2017} found that the vast majority of backward compatibility issues are solved in the app code. For this reason, when a new version of a mobile operating system is released, app developers are concerned with its impact on their apps. 

Migrating an app to newer APIs is a tedious and challenging task. Once an incompatibility is discovered, developers have to analyze the behavior of the app to identify the cause of the problem and the location of the fix, to then implement and validate the fix. The whole process can be demanding and makes developers reluctant to migrate to newer APIs. For instance, McDonnell et al.~\cite{McDonnell:APIStability:ICSM:2013} report an average migration time of 16 months, in contrast to an average API release interval of 2 months~\cite{AndroidVersionHistory}.

So far, several approaches have been proposed to identify the problems introduced during framework upgrades~\cite{Li:CompatibilityIssues:ISSTA:2018,Mostafa:BackwardIncompatibilities:ISSTA:2017,Mora:ClientSpecificEquivalenceChecking:ASE:2018,Fazzini:DiffDroid:ASE:2017}, but little effort has been directed towards supporting their localization and resolution. Spectrum-based fault localization techniques (SBFL)~\cite{Wong:SurveyLocalization:TSE:2016} can be used to localize problems, however they are not effective in this scenario, as reported in our evaluation, and require a test suite with both failing and passing test cases, which is not always available for mobile apps. Anomaly detection techniques can be an alternative option to identify suspicious behaviors, but they also require extensive test suites of passing test cases to infer models and do not offer any localization capability~\cite{Mariani:BCT:TSE:2011,Pastore:Radar:ISSRE:2012,Pradel:SpecMining:ICSE:2012,Zuddas:Mimic:ASE:2014}. 

This paper describes a tool that implements the FILO\linebreak 
technique~\cite{Mobilio:FILO:ISSRE:2019}, which is a technique specifically designed to facilitate developers in resolving backward compatibility issues introduced by Android upgrades. FILO, by relying on one failing test case only, automatically identifies the methods of the app that likely need to be modified to fix an issue caused by the upgrade of the Android framework. FILO also reports the key symptomatic anomalous events observed in the failed execution that can explain and facilitate the understanding of the problem that must be fixed.


The tool was tested on 12 real-world backward incompatibilities reported on GitHub and evaluated by a small group of third-party Android developers who provided qualitative feedback about the approach and the tool. Results indicate that FILO can actually help developers providing relevant recommendations and distilling useful information.  

The paper is organized as follows. Section~\ref{sec:filo} describes the FILO technique, its architecture and usage. Section~\ref{sec:empiricalResults} reports empirical results obtained using the tool. Finally, Section~\ref{sec:conclusions} provides concluding remarks. 


\section{FILO}
\label{sec:filo}
This section describes the FILO technique, the architecture of the tool that implements the technique, and a usage example.

\subsection{Technique}
\label{subsec:technique}
Given an Android app affected by a backward incompatibility issue, FILO automatically  generates a ranked list of methods corresponding to the possible fix locations. Each method in the ranked list is associated with a set of suspicious invocations produced by the interaction between the app and the underlying framework, which may help developers understanding both \emph{why} the method may contain code incompatibile with the upgraded version of the framework and \emph{how} to fix the app. 
FILO needs three inputs to produce this information: one automatic test case that reveals the backward incompatibility problem caused by a framework upgrade and two Android environments to run the test, one running the version of API that makes the test case pass, and the other running the version of the API that makes the test case fail.  Note that FILO, unlike SBLF techniques, does not depend on the coverage of an input test suite, as it requires only one test case that reveals the incompatibility problem. FILO produces a ranked list of methods representing the possible fix locations as output. Each entry in the list is enriched with invocations observed during the test that may justify the identification of the method as a possible fix locus.

FILO runs in three phases: (1) \emph{Test Execution}, which runs the available test case in both the environments, collecting traces about the interactions (i.e., API calls and callbacks) between the app and the frameworks; (2) \emph{Anomaly Detection}, which analyzes the collected traces to reveal differences that may explain the failure;  and (3) \emph{Fix Locus Candidates Identification}, which exploits the differences to derive a ranked list of app methods that constitute the fix locus candidates, each of them enriched with the suspicious differences that motivated their ranking. Note that, while suspicious differences are limited to the interactions between the app and the framework, the fix locus candidates can be any method of the app.

\subsubsection*{Test Execution} In this phase, the available test case is executed within the two Android environments and the interactions between the app and the framework are recorded. These interactions include all the calls to framework methods originated from the app (API calls), and all the calls to app methods originated by the framework (callbacks), while internal calls (calls between methods of the app or between methods of the framework) are not recorded. Since the analyzed problem is a backward incompatibility introduced by a framework upgrade, the intuition is that clear evidence about the symptoms of the failure should be observable from the comparison of the interactions between the app and the framework when the same test is executed in the two environments. We refer to the trace obtained from the passing test as the \emph{baseline trace}, while we refer to the other trace as the \emph{failure trace}. Note that the recorded traces include method names, parameters, and return values.


\subsubsection*{Anomaly Detection}
The anomaly detection phase uses \textit{diff}~\cite{LinuxDiff} to compare the execution traces resulting from the previous phase to identify a set of invocation blocks that look suspicious. An invocation block is defined as a set of contiguous interactions between the Android framework and the app that are extracted from the traces. The differences between the traces are considered suspicious since they are caused by the upgraded execution environment and therefore likely related to the observed backward incompatibility. For this reason we refer to the blocks returned by diff as the \textit{Suspicious Invocation Blocks (SIBs)}. FILO heuristically weights each SIB according to the number of boundary calls that it contains (i.e., a block with 3 methods calls has weight 3) because irrelevant differences often consist of a single or few different method calls, while relevant differences often produce entire new behaviours that cause the execution of several method calls. The set of weighted SIBs are used in the last phase to recommend the locations where the app should be modified to correctly interact with the upgraded framework. 


\subsubsection*{Fix Locus Candidates Identification}

Since SIBs represent the symptoms, but not the source of the problem, FILO also considers the methods invoked before reaching the methods in the SIBs as possible locations for the fault. In particular, FILO builds a \textit{failure call tree} that represents the methods executed during the failure and their call-callee relationship. The SIBs are the leaves of the tree, the methods that invoke the first method of the SIBs are their direct ancestors, and so on transitively until the root of the tree that represents the \texttt{ZygoteInit.main} method, which is the root of every Android program.


Each node (i.e., method) in the tree is heuristically weighted based on its likelihood to represent the location where the app should be modified to fix the observed problem. The weight of a node depends on two factors: the (weighted) number of SIBs that can be reached from the node and the position of the method in the hierarchy. The first factor privileges the nodes that occur higher in the hierarchy (e.g., the root node has the maximum score for this factor since it is possible to reach every SIB from the root of the tree), while the second factor prioritizes the nodes occurring close to the leaves (e.g., the leaves have maximum score for this factor). The intuition is that starting from leaves, it is better to consider a method occurring higher in the hierarchy as possible locus of the fix only if it originates a significative number of relevant SIBs. The interested reader can access~\cite{Mobilio:FILO:ISSRE:2019} for the details about the scoring process and formula. Finally, the SIBs that can be reached from each node are used as supporting evidence of the symptoms that should be fixed by modifying the method represented by the node. This information can be useful to understand why and how a given method should be modified.

\subsection{Tool Architecture}
\label{subsec:architecture}

\begin{figure}[tb]
  \centering
  \includegraphics[width=0.5\textwidth] {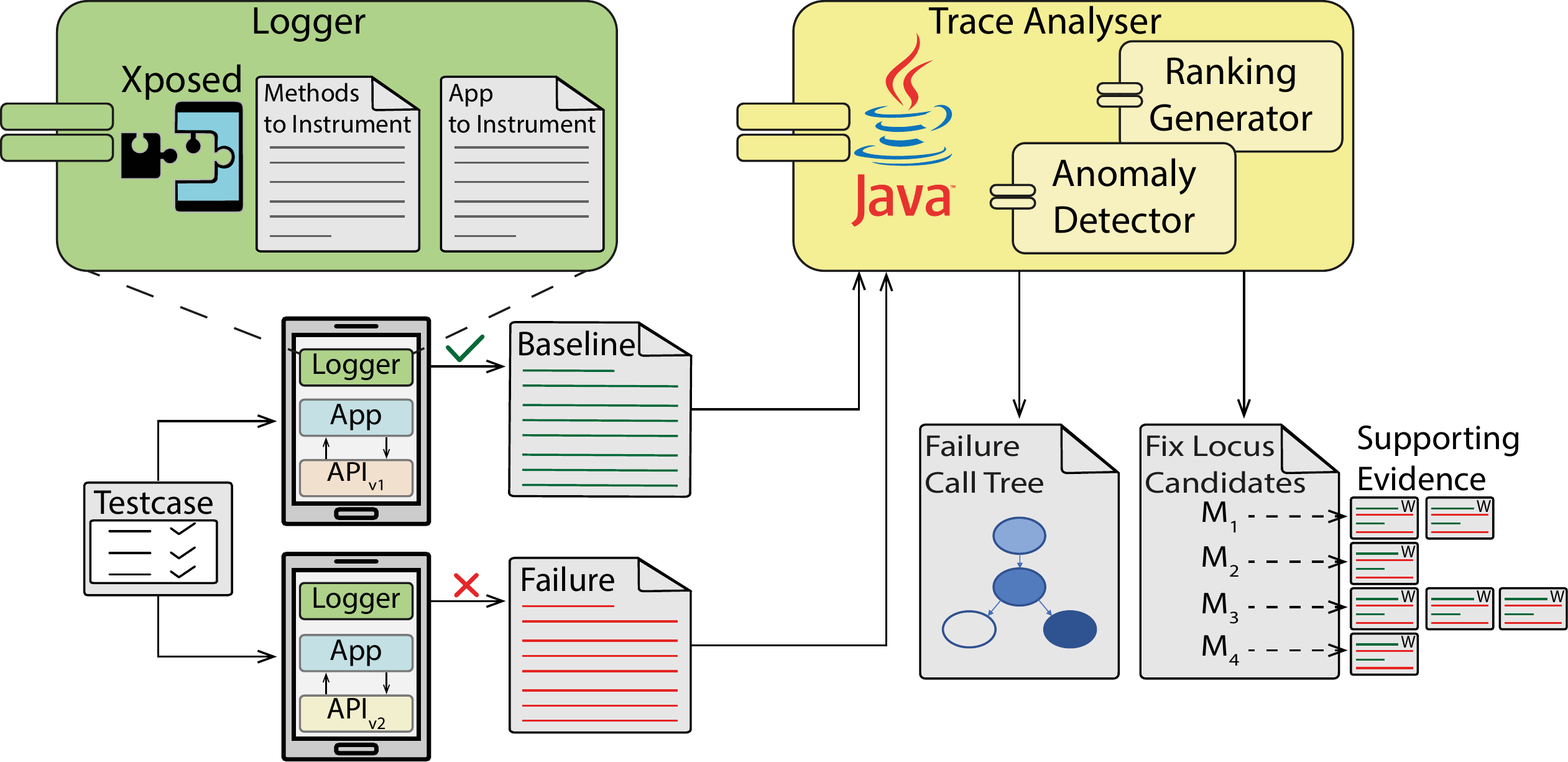}

  \caption{Overview of FILO components.}
  \label{fig:filoOverview}
\end{figure}

The tool consists of two main components: the \texttt{FILO Logger}, which is executed in the device and is in charge of collecting the \textit{baseline} and \textit{failure trace}s, and the \texttt{FILO Trace Analyser}, which is executed outside the device and is in charge of analizing the traces to produce the ranking and the supporting evidence. The architecture of the tool is shown in Figure~\ref{fig:filoOverview}.

\subsubsection*{FILO Logger}
The \texttt{FILO Logger} is implemented as an Xposed module. Xposed is an opensource framework that can be used to dynamically inject functionalities into Android apps when loaded for execution, similarly to Aspect-oriented frameworks~\cite{Xposed_2016}. Our Xposed module is able to trace the interactions between an app and its framework, including parameter and return values. Since instrumenting every method can be prohibitive, FILO selectively instruments only the methods that may occur as boundary calls of the analyzed test case (namely the \textit{Methods to instrument}). To obtain this list of methods we first run the test case with the lightweight Android tracer, which inexpensively produces the list of all methods executed with both versions of the framework. We then post-process this list to automatically obtain the methods that must be monitored by our logger to produce the information required by FILO. To filter the list of methods, the \texttt{FILO Logger} uses the app identifier (namely the \textit{App to instrument}), that is the app's root package name.



\subsubsection*{FILO Trace Analyser}
The \texttt{FILO Trace Analyser} includes two subcomponents, the \texttt{Anomaly Detector} and the \texttt{Ranking} \texttt{Generator}. The \texttt{Anomaly Detector} comparers the traces, and identifies the SIBs, along with their weights. The \texttt{Ranking Generator} builds the failure call tree, runs the scoring algorithm, and produces the ranking, which also includes the supporting evidence. The \texttt{FILO Trace Analyser} is implemented in Java. The failure call tree is outputted as a DOT file, while the ranking and supporting evidence are generated as CSV file. The latter file contains the ordered list of methods, each of which is associated with a set of suspicious invocations representing symptoms of failure that can be removed by implementing an appropriate fix in the method.  

\smallskip

All the steps of the approach can be executed from scripts, while outputs use standard formats, easing the integration of the tool within continuous testing environments. In particular, at the current state of development, FILO consists of a trace recorder for Android applications and a command line tool for the analysis of the execution traces. Furthermore we also provide a command line tool that extracts the set of methods to be instrumented from the traces saved by the Android tracer and outputs this information complying with the input format required by FILO.

\subsection{Usage Example}
\label{subsec:usecase}
This section presents how a developer should use FILO to analyze an app affected by a backward compatibility issue. As a sample app we refer to Good Weather\footnote{https://play.google.com/store/apps/details?id=org.asdtm.goodweather}, which suffers from an upgrade problem due to a breaking change introduced in API 23\footnote{https://github.com/qqq3/good-weather/issues/3}. Because of this problem, the app runs correctly with API 22, but it hangs showing a loading spinner when running with API 23.


The developer can analyze this problem by recording an automatic test case (e.g., with Appium~\cite{Appium}) and running the test case with the Android tracer enabled. 
The resulting trace is processed using the script we developed and made available to obtain the list of methods to be instrumented.
The developer can thus analyze the failure by running the \texttt{FILO Logger} on both API 22 and API 23 specifying Good Weather as the app to instrument and passing the generated list of methods to the tool. This execution produces the \emph{baseline trace} and the \emph{failure trace}. These traces are finally analyzed by running the \texttt{Trace Analyser} that first executes the \texttt{Anomaly Detector}, which detects the SIBs, and then the \texttt{Ranking Generator}, which produces the ranking and the supporting evidence.

In the case of Good Weather, the ranking generator places the\linebreak {\footnotesize \texttt{gpsRequestLocation}} and {\footnotesize \texttt{onOptionsItemSelected}} methods at the top of the ranking, suggesting that the problem might be due to the access to the GPS location. Interestingly, the supporting evidence associated with these methods include the {\footnotesize \texttt{checkSelfPermission}} and {\footnotesize \texttt{requestLocationUpdates}} methods, which suggest that there might be a problem with the permission to access location data. These suggestions are correct, since the framework upgrade changed the policy to access resources, forcing apps to make an explicit permission request when they access a resource for the first time. Adding this request in the methods at the top of the ranking is exactly the way the fix is implemented in the app\footnote{https://github.com/qqq3/good-weather/commit/81eab554bae5299e33a4ce9babb690359647115c}, as well pointed out by FILO. 
%

\section{Empirical Results}
\label{sec:empiricalResults}
In our previous work~\cite{Mobilio:FILO:ISSRE:2019}, we evaluated FILO with 12 actual backward compatibility issues reported on GitHub. The evaluation focused on assessing: $(i)$ \emph{the relevance of the captured information}, $(ii)$ \emph{the quality of the ranking}, and $(iii)$ the effectiveness of FILO in \emph{comparison} to \emph{SBFL} and \emph{naive trace analysis}. 
We consider Ochiai~\cite{Abreu:2007} as representative of SBFL techniques, while naive trace analysis represents the straightforward approach of inspecting the differences between the baseline and failure traces in the order they occur. In addition to summarizing these empirical results, we report novel, although preliminary, qualitative results  about the usefulness of the tool as $(iv)$ \emph{perceived by Android developers}.

\textbf{\textit{(i)} Relevance of the captured information.} To evaluate this aspect, we measured the relevance of the information present in the SIBs, which are the basic building blocks of the analysis implemented by FILO. To objectively determine the interactions that occur as direct consequence of the backward incompatibility introduced by the framework upgrade, we executed the version of the app where the problem was fixed and collected the same trace collected for the faulty app. The interactions between the app and the framework that changed due to the fix are used as the set of the relevant interactions that should be captured by the SIBs. We discovered that the percentage of SIBs containing relevant information ranged in the interval 50\%-100\%, confirming their suitability for the analysis. 

\textbf{\textit{(ii)} Quality of the ranking.} To assess the quality of the ranking returned by FILO we considered the position of the method that has been modified to achieve the fix. Interestingly, FILO was always able to report the method to be fixed within the first 10 positions, except for an app where the method was not part of the failure call tree. In 7 cases the method to be changed was ranked among the top 5 positions with 4 perfect results, that is, the method to be changed was ranked first. This is in line with the preference of practitioners who consider inspection of up to 10 methods acceptable, with a preference for techniques that require the inspection of up to 5 methods~\cite{Kochhar:LocalizationExpectation:ISSTA:2016}.

\textbf{\textit{(iii)} Comparison to SBFL and Naive Trace Analysis.} We compared FILO to naive trace analysis, which is our baseline method, and Ochiai~\cite{Abreu:2007} executed both at the granularity of methods and statements. Since a passing test suite is needed to run SBFL techniques and no test suite was available for the considered apps, we generated passing test cases with Monkey~\cite{Monkey}.

The results of the comparison are shown in Table~\ref{tab:comparison}.  For each row, we have the symbol ``-'' if the method is not found and the best result is shown in bold. Rows \emph{Top-1}, \emph{Top-5}, and \emph{Top-10} indicate the number of times each technique has ranked the method to be fixed in the top 1, top 5, and top 10 positions, respectively. The \emph{Not in the ranking} row shows the number of times that a technique has not included the method to be fixed in the ranking. 

FILO has been always more effective than naive trace analysis. Ochiai is better than FILO in only two cases (2 out of 12), but in the remaining cases FILO has always ranked the method to be fixed significantly better than Ochiai, confirming the effectiveness of FILO in the localizaton of the problems introduced by framework upgrades.

\begin{table}[th]
\caption{Comparison between methods.} \label{tab:comparison}\vspace{-0.5cm}
\begin{footnotesize}
\begin{center}
\begin{tabular}{@{}lc c c c@{}}
\toprule
Application  & FILO & Naive Trace & Ochiai & Ochiai \\
  &  &  Analysis & (method) & (statement) \\
\midrule

BossTransfer &  \textbf{2} & 138 & 4 & 32\\

FakeGPS      & \textbf{5} & 328 &  13 & 65\\

FilePicker   & \textbf{4} & 203 & 81 & -\\

GetBack GPS  & \textbf{10} & - & - & - \\

GoodWeather  & \textbf{1,2} & - & 1,32 & 5,5 \\

KanjiFix     & \textbf{1} & - & 19  & 23\\

MapDemo      & 8 & 45 & \textbf{1} & \textbf{1}\\

PoGoIV & \textbf{7} & - & 48 & 283  \\ 

PrivacyPolice & \textbf{1} & 26 & 21 &130 \\

QuotoGraph & \textbf{1} & - & - &- \\

SearchView   & \textbf{9} & 109 & - & -\\

ToneDef   & - & - & 24 &  \textbf{13}\\ \midrule

Top-1 & \textbf{4} & 0 & 2 & 1\\
Top-5 & \textbf{7} & 0 & 3 & 2 \\
Top-10 & \textbf{11} & 0 & 3 & 2 \\
Not in the ranking & \textbf{1} & 6 & 3 & 4\\

\bottomrule
\end{tabular}
\end{center}
\end{footnotesize}
\end{table}

\textbf{\textit{(iv)} Usefulness to Android developer.} To investigate if the FILO tool can practically support developers in solving problems caused by Android upgrades, we involved 4 junior third-party Android developers in the task of locating and fixing the faults in the PrivacyPolicy and PoGoIV apps. The developers were provided with the source code of the two apps and the test that reproduces the failure, but they received the ranking only for one of the apps. In particular, two developers received the ranking for PrivacyPolicy but not for PoGoIV, while the other two developers worked with the opposite setting. Developers were also introduced to the functionalities implemented by the apps.

Results show that when provided with the ranking produced by FILO, the subjects were able to localize the faults in a few minutes, while the faults have been localized only 50\% of the times when the ranking was not available. We also submitted an exit questionnaire asking if the ranking and the supporting evidence were useful, non-influential, or misleading. The output of the tool has never been considered misleading, and 3 out of the 4 subjects considered it useful, while one subject considered it non-influential. This preliminary qualitative evidence increases the confidence on the practical utility of the tool.



Our preliminary evaluation with human subjects shows that the ranking can be useful to developers with little familiarity with the code of the faulty app. However, some faults can be non-trivial to fix even for the developers of the apps. For instance, the faults that we used in our evaluation have been released after a period that ranged between 1 day and several months. This suggests that debugging aids, such as the one offered by FILO, might be useful also for developers who are familiar with the codebase of the application under test.

\section{Conclusions}
\label{sec:conclusions}
The Android framework is regularly updated~\cite{Wei:AndroidFragmentation:ASE:2016,Li:DeprecatedAPI:MSR:2018,AndroidVersionHistory} introducing backward incompatible changes that cause misbehaviors in the existing apps~\cite{Wei:AndroidFragmentation:ASE:2016,Mostafa:BackwardIncompatibilities:ISSTA:2017}. In order to fix these problems, developers have to spend effort analyzing the behavior of the apps to identify the cause of the problems and the location of the fixes.

This paper presents FILO, a tool that helps developers in the identification of the methods that must be modified to make a failing app compatible with a newer version of the Android environment. The suggested methods are arranged in a ranked list and are enriched with additional information that help developers interpreting the ranking, easing the implementation of the correct fix.
Our evaluation shows that the tool can be practically useful to Android developers.

\begin{small}
\subsubsection*{Acknowledgements}
This work has been partially supported by the H2020 ERC CoG Learn project (grant agreement n. 646867).
\end{small}

\vfill

\bibliographystyle{ACM-Reference-Format}
\bibliography{Main.bib}

\end{document}